\documentclass[3p,times,twocolumn]{elsarticle}

\usepackage{ecrc}

\volume{00}

\firstpage{1}

\journalname{Nuclear Physics B Proceedings Supplement}

\runauth{}

\jid{nuphbp}

\jnltitlelogo{Nuclear Physics B Proceedings Supplement}

\usepackage{amssymb}
\usepackage{amsmath}
\usepackage{amsthm}

\usepackage[figuresright]{rotating}

\def\alf1{ {\alpha\over\pi} }
\def\rQCED{{\rm QCED}}
\begin{document}

\begin{frontmatter}



\dochead{\small\bf BU-HEPP-14-05, Aug., 2014}

\title{Comparisons of Exact Amplitude--Based Resummation Predictions and LHC Data }


\author[label1]{B.F.L. Ward}
\author[label2]{S.K. Majhi}
\author[label1]{A. Mukhopadhyay}
\author[label3]{S.A. Yost}

\address[label1]{Baylor University, Waco, TX, USA}
\address[label2]{IACS, Calcutta, IN}
\address[label3]{The Citadel, Charleston, SC, USA}

\begin{abstract}
Using the MC  Herwiri1.031, we present the current status of the comparisons with LHC data of the predictions of our approach of exact amplitude-based resummation for precision QCD calculations. 
\end{abstract}

\begin{keyword}
QCD Resummation IR-Improved DGLAP-CS Theory NLO-PS MC

\end{keyword}

\end{frontmatter}


\def\Kmax{K_{\rm max}}\def\ieps{{i\epsilon}}\def\rQCD{{\rm QCD}}
\section{\bf Introduction}\label{intro}\par
The successful running of the LHC  during 2010-2012 has resulted in large data samples on SM standard candle processes such as heavy gauge boson production and decay to lepton pairs (samples exceeding $10^7$ of such
events for $Z/\gamma^*$ production) for ATLAS and CMS.
Such data signal the arrival of the era of precision QCD, with
predictions for QCD processes at the total precision tag of $1\%$ or better, and make more manifest the need for exact, 
amplitude-based resummation of large higher order effects as discussed in Refs.~\cite{herwiri1}.  
Such precision allows
one to  
distinguish new physics(NP) from higher order SM processes and to distinguish 
different models of new physics from one another as well. 
We present here comparisons of the attendant 
application of exact amplitude-based
resummation theory to recent data
from the LHC.
We first review the elements our approach as formulated in Ref.~\cite{qced} before we turn 
in the next section to comparisons with recent LHC data.\par 
Our starting point is the
well-known representation
\begin{equation}
d\sigma =\sum_{i,j}\int dx_1dx_2F_i(x_1)F_j(x_2)d\hat\sigma_{\text{res}}(x_1x_2s)
\label{bscfrla}
\end{equation}
of a hard LHC scattering process, where $\{F_j\}$ and 
$d\hat\sigma_{\text{res}}$ are the respective parton densities (PDF's) and 
reduced resummed hard differential cross section. The resummation includes 
all large EW and QCD higher order corrections as needed
for achieving a total precision tag of 1\% or better for the total 
theoretical precision of (\ref{bscfrla}). 
The total theoretical precision $\Delta\sigma_{\text{th}}$ of (\ref{bscfrla}) as defined in Refs.~\cite{jadach-prec,radcr11}  
is essential to the faithful application
of any theoretical prediction to precision experimental data.
Whenever $\Delta\sigma_{\text{th}}\leq f\Delta\sigma_{\text{expt}}$, 
where $\Delta\sigma_{\text{expt}}$ is the respective experimental error
and $f\lesssim \frac{1}{2}$,
the theoretical uncertainty will not adversely affect the physics
analysis of the data. 
With our eye on a provable theoretical precision tag  
we have 
developed the $\text{QCD}\otimes\text{QED}$ resummation theory in Refs.~\cite{qced}
for (\ref{bscfrla}).
The key exact master formula is  
{\small
\begin{eqnarray}
&d\bar\sigma_{\rm res} = e^{\rm SUM_{IR}(QCED)}
   \sum_{{n,m}=0}^\infty\frac{1}{n!m!}\int\prod_{j_1=1}^n\frac{d^3k_{j_1}}{k_{j_1}} \cr
&\prod_{j_2=1}^m\frac{d^3{k'}_{j_2}}{{k'}_{j_2}}
\int\frac{d^4y}{(2\pi)^4}e^{iy\cdot(p_1+q_1-p_2-q_2-\sum k_{j_1}-\sum {k'}_{j_2})+
D_\rQCED} \cr
&\tilde{\bar\beta}_{n,m}(k_1,\ldots,k_n;k'_1,\ldots,k'_m)\frac{d^3p_2}{p_2^{\,0}}\frac{d^3q_2}{q_2^{\,0}}.
\label{subp15b}
\end{eqnarray}}\noindent
Here $d\bar\sigma_{\rm res}$ is either the reduced cross section
$d\hat\sigma_{\rm res}$ or the differential rate associated to a
DGLAP-CS~\cite{dglap,cs} kernel involved in the PDF evolution and 
the {\em new} (YFS-style~\cite{yfs,sjbw}) {\em non-Abelian} residuals 
$\tilde{\bar\beta}_{n,m}(k_1,\ldots,k_n;k'_1,\ldots,k'_m)$ have $n$ hard gluons and $m$ hard photons and we show the generic $2f$ final state 
with momenta $p_2,\; q_2$ for
definiteness. The infrared functions ${\rm SUM_{IR}(QCED)}$, $ D_\rQCED\; $
are given in Refs.~\cite{qced,irdglap1,irdglap2}. The residuals $\tilde{\bar\beta}_{n,m}$ allow a rigorous parton 
shower/ME matching via their shower-subtracted 
counterparts $\hat{\tilde{\bar\beta}}_{n,m}$~\cite{qced}.\par
We now discuss the paradigm opened by (\ref{subp15b}) for precision QCD via comparisons with recent data.\par

\section{Comparisons to Data of Precision QCD for the LHC}
We first recall that, as we have discussed in Refs.~\cite{herwiri1}, the methods we employ 
are fully consistent with the methods in Refs.~\cite{stercattrent1,scet1,css,resbos,banfi,neubrt} but we do not have intrinsic physical physical barriers to sub-1%
precision as do the approaches used in the latter references. They may used to give approximations to our new residuals $\tilde{\bar\beta}_{m,n}$ for studies of consistency~\cite{elswh}. 
\par
With this understanding, we note that, if we apply 
(\ref{subp15b}) to the
calculation of the kernels, $P_{AB}$, we arrive at 
an improved IR limit of these kernels, IR-improved DGLAP-CS theory . In this latter theory~\cite{irdglap1,irdglap2} large IR effects are resummed for the kernels themselves.
From the resulting new resummed kernels, $P^{exp}_{AB}$~\cite{irdglap1,irdglap2} we get a new resummed scheme for the PDF's and the reduced cross section: 
$F_j,\; \hat\sigma \rightarrow F'_j,\; \hat\sigma'\; \text{for}$
$P_{gq}(z)\rightarrow P^{\text{exp}}_{gq}(z)=C_FF_{YFS}(\gamma_q)e^{\frac{1}{2}\delta_q}\frac{1+(1-z)^2}{z}z^{\gamma_q}, \text{etc.}.$
This new scheme gives $\sigma$ in (\ref{bscfrla}) with improved MC stability~\cite{herwiri1}. 
Here, $C_F$ is the quadratic Casimir invariant for the quark color representation.
See Refs.~\cite{irdglap1,irdglap2} for the definitions of $F_{YFS},\gamma_q,\; \delta_q$ as well as for the complete
set of results for the new $P^{exp}_{AB}$.\par
The physical idea underlying the new kernels was shown by Bloch and Nordsieck~\cite{bn1}: 
due to the coherent state of very soft massless gauge field 
quanta generated by an accelerated charge it impossible to know which of the infinity of possible states
one has made in the splitting process $q(1)\rightarrow q(1-z)+G\otimes G_1\cdots\otimes G_\ell,\; \ell=0,\cdots,\infty$.
The new kernels take this effect into account by resumming the terms
${\cal O}((\alpha_s\ln(q^2/\Lambda^2)\ln(1-z))^n)$ for the IR limit $z\rightarrow 1$. This resummation generates~\cite{herwiri1,irdglap1,irdglap2} the Gribov-Lipatov exponents $\gamma_A$ which start in ${\cal O}(\hbar)$ in the loop expansion~\footnote{See Ref.~\cite{ope-dglap} for the 
connection between the new kernels and the Wilson expansion.}.\par
\begin{figure}[h]
\begin{center}
\includegraphics[width=80mm]{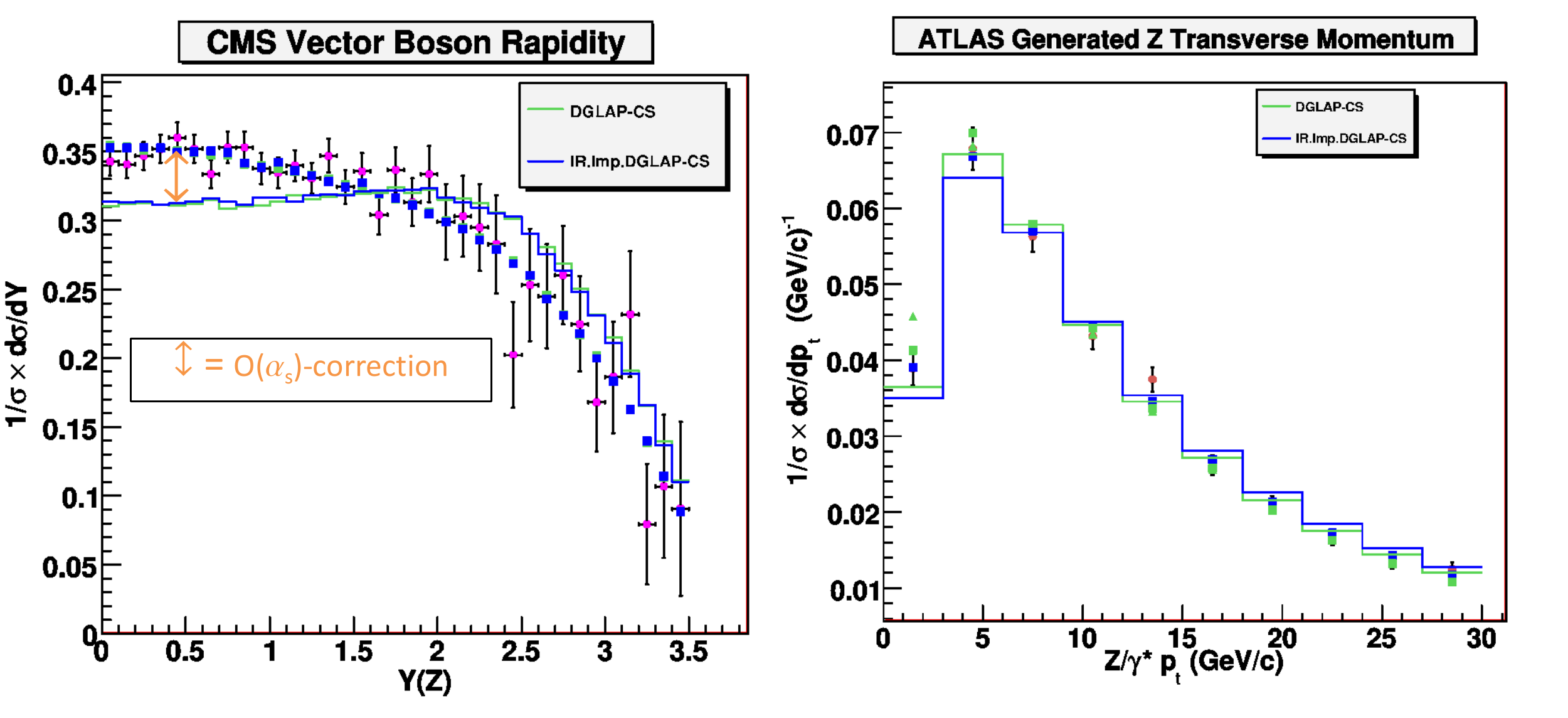}
\end{center}
\caption{\baselineskip=8pt Comparison with LHC data: (a), CMS rapidity data on
($Z/\gamma^*$) production to $e^+e^-,\;\mu^+\mu^-$ pairs, the circular dots are the data, the green(blue) lines are HERWIG6.510(HERWIRI1.031); 
(b), ATLAS $p_T$ spectrum data on ($Z/\gamma^*$) production to (bare) $e^+e^-$ pairs,
the circular dots are the data, the blue(green) lines are HERWIRI1.031(HERWIG6.510). In both (a) and (b) the blue(green) squares are MC@NLO/HERWIRI1.031(HERWIG6.510($\rm{PTRMS}=2.2$GeV)). In (b), the green triangles are MC@NLO/HERWIG6.510($\rm{PTRMS}=$0). These are otherwise untuned theoretical results. 
}
\label{fig2-nlo-iri}
\end{figure}

 The first realization of the new IR-improved kernels is given by new MC Herwiri1.031~\cite{herwiri1} in the Herwig6.5~\cite{hrwg} environment. Realization of the new kernels in the Herwig++~\cite{hwg++}, Pythia8~\cite{pyth8}, Sherpa~\cite{shrpa} and Powheg~\cite{pwhg} environments is in progress as well.  In Fig.~\ref{fig2-nlo-iri} we illustrate some of the recent comparisons we have made between Herwiri1.031 and Herwig6.510, 
both with and without
the MC@NLO~\cite{mcnlo} exact ${\cal O}(\alpha_s)$ correction\footnote{See Refs.~\cite{herwiri1} for the connection bewteen the $\hat{\tilde{\bar\beta}}_{n,m}$ and the MC@NLO differential cross sections.} , 
in relation to the LHC data~\cite{cmsrap,atlaspt} on $Z/\gamma*$ production with decay to lepton pairs\footnote{Similar comparisons were made in relation to such data~\cite{d0pt,galea} from FNAL 
in Refs.~\cite{herwiri1}.}.
Just as we found in Refs.~\cite{herwiri1} for the FNAL data on single $Z/\gamma^*$ production, the unimproved MC requires the very hard value of $\text{\rm PTRMS}\cong 2.2$GeV 
to give a good fit to the $p_T$ spectra as well as the rapidity spectra whereas the IR-improved calculation gives very good fits to both of the spectra without the need of such a hard value of {\rm PTRMS}, the rms value for
an intrinsic Gaussian $p_T$ distribution, for the proton wave function: the $\chi^2/d.o.f$ are respectively $(0.72,0.72),\;(1.37,0.70),\;
(2.23,0.70)$ for the $p_T$ and rapidity data for the MC@NLO/HERWIRI1.031,
 MC@NLO/HERWIG6.510($\rm{PTRMS}=2.2$GeV) and MC@NLO/HERWIG6.510($\rm{PTRMS}=$0)
results. Such a hard intrinsic value of {\rm PTRMS} contradicts the results
in Refs.~\cite{rvndl,bj}, as we discuss in Refs.~\cite{herwiri1}.
To illustrate the size of the exact ${\cal O}(\alpha_s)$ correction, we also show the
results for both Herwig6.510(green line) and Herwiri1.031(blue line) without it in the plots in Fig.~\ref{fig2-nlo-iri}. As expected, the exact ${\cal O}(\alpha_s)$ correction is important for both the $p_T$ spectra and the rapidity spectra. The suggested accuracy at the 10\% level shows
the need for the NNLO extension of MC@NLO, in view of our goals
for this process. We also note that, with the 1\% precision goal, one also needs per mille level control of the EW corrections. This issue is addressed in the new version of the ${\cal KK}$ MC~\cite{kkmc422}, version 4.22, which now allows for incoming quark antiquark beams -- see Ref.~\cite{kkmc422} for further discussion of the relevant effects in relation 
to other approaches~\cite{otherEW}.\par
\begin{figure}[h]
\begin{center}
\setlength{\unitlength}{0.1mm}
\begin{picture}(800, 465)
\put( 175, 400){\makebox(0,0)[cb]{\bf (a)} }
\put(600, 400){\makebox(0,0)[cb]{\bf (b)} }
\put(   -50, 0){\makebox(0,0)[lb]{\includegraphics[width=40mm]{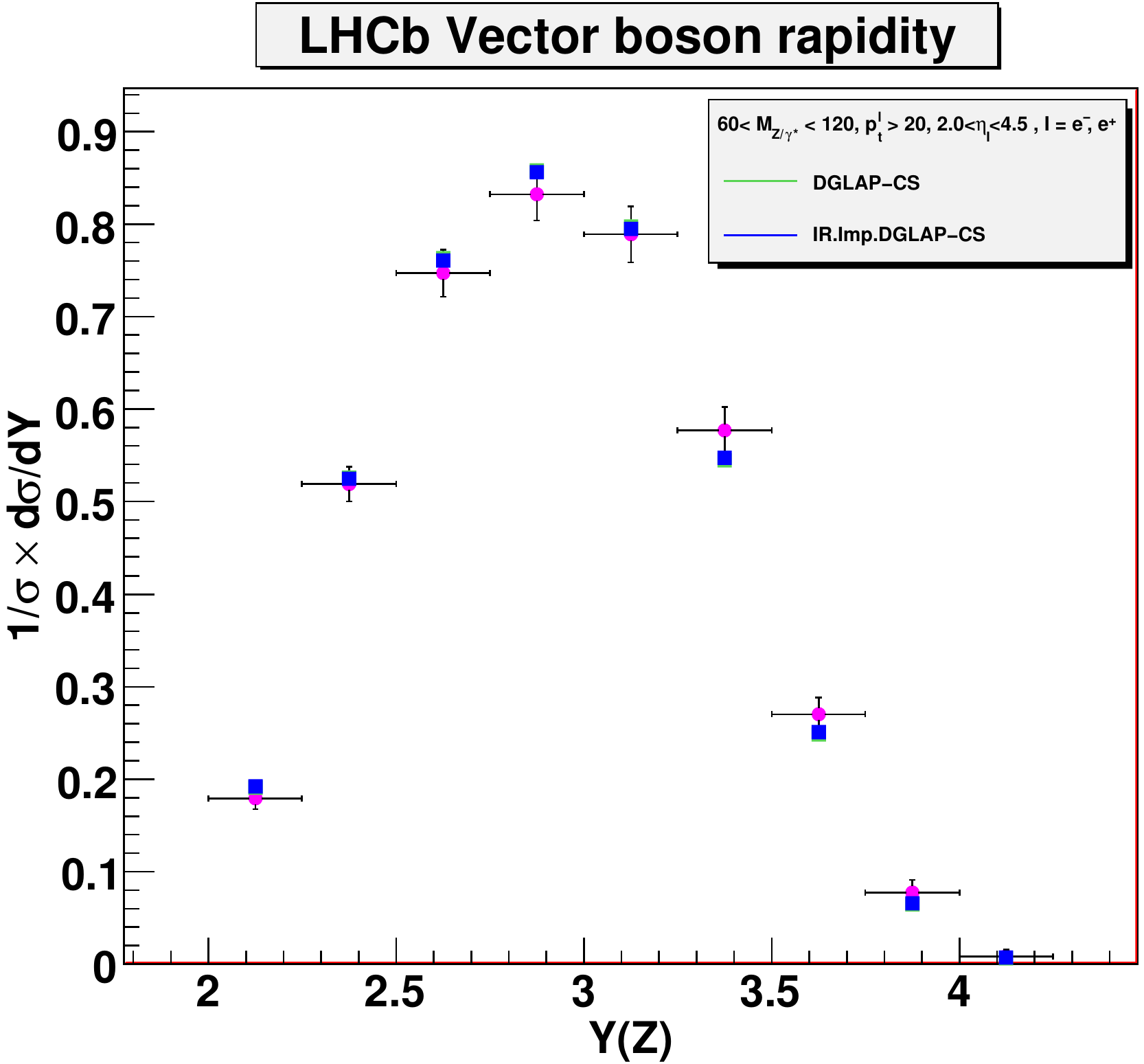}}}
\put( 390, 0){\makebox(0,0)[lb]{\includegraphics[width=40mm]{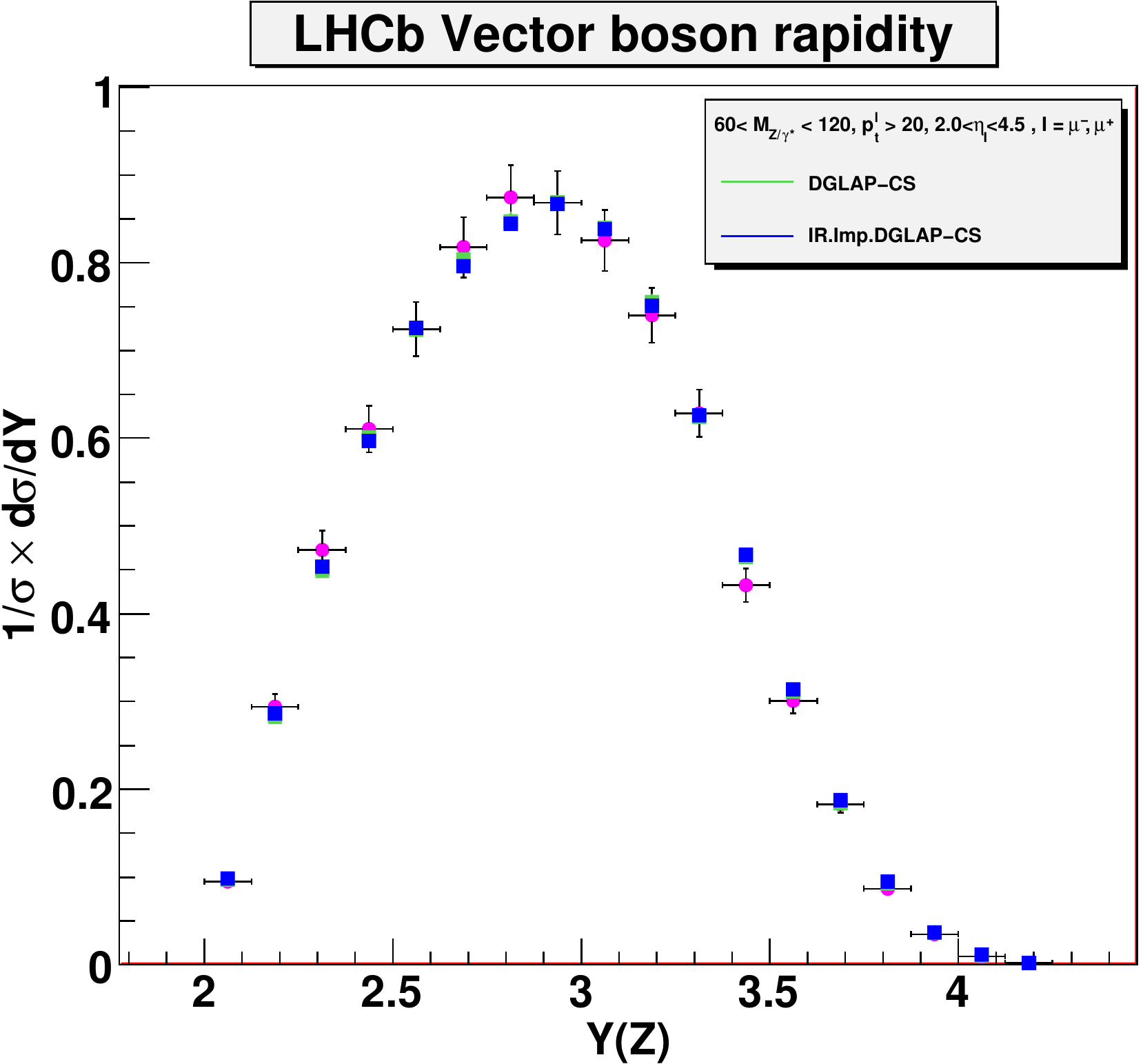}}}
\end{picture}
\end{center}
\caption{\baselineskip=8pt Comparison with LHC data: LHCb rapidity data on
($Z/\gamma^*$) production to (a) $e^+e^-,\;(b) \;\mu^+\mu^-$ pairs, the circular dots are the data. In both (a) and (b) the blue(green) squares are MC@NLO/HERWIRI1.031(HERWIG6.510($\rm{PTRMS}=2.2$GeV)) and the green triangles are MC@NLO/HERWIG6.510($\rm{PTRMS}=$0). These are otherwise untuned theoretical results. 
}
\label{fig-lhcb1}
\end{figure}

 
We have also made comparisons with recent LHCb data~\cite{lhcbdata} on single $Z/\gamma^*$ production and decay to lepton pairs. These results will be presented in detail
elsewhere~\cite{elswh}. Here, we illustrate them with the results in Fig.~\ref{fig-lhcb1} for the $Z/\gamma^*$ rapidity as measured by LHCb for the decays to $e^+e^-$ pairs and the decays to $\mu^+\mu^-$ pairs. 
These data probe a different phase space regime: the 
lepton pseudorapidity $\eta$ satisfies $2.0<\eta<4.5$ 
to be compared with  
$|\eta_\ell|<2.1$($|\eta_\ell|<2.4$), $|\eta_{\ell'}|<4.6$($|\eta_{\ell'}|<2.4$) for the CMS(ATLAS) data in Fig.~\ref{fig2-nlo-iri}. Here $\eta_\ell(\eta_{\ell'})$ is the respective pseudorapidity of $\ell,\; \ell=\mu,\bar{\mu}(\ell',\; \ell'=e,\bar{e}),$ respectively. Again, the agreement between the IR-improved MC@NLO/Herewiri1.031 without the need of an 
ad hocly hard value of ${\rm PTRMS}$ is shown for  both the $e\bar{e}$ and $\mu\bar{\mu}$ data, where the $\chi^2/d.o.f.$ are $0.746, 0.773$ respectively. The unimproved calculations with MC@NLO/Herwig6510
for ${\rm PTRMS}=0$ and ${\rm PTRMS}=2.2$ GeV 
respectively also give good fits, with the $\chi^2/d.o.f.$ of $0.814,\;0.836$ and $0.555,\;0.537$ respectively for the $e\bar{e}$ and $\mu\bar{\mu}$ data.
In the phase space probed by the LHCb, it continues to hold that the more inclusive observables such as the normalized $Z/\gamma^*$ rapidity spectrum are not as sensitive to the IR-improvement as observables such as the $Z/\gamma^*$ $p_T$ spectrum. \par
As one has now more than $10^7$ $Z/\gamma*$ decays to lepton pairs per experiment at ATLAS and CMS, we show in Refs.~\cite{herwiri1} that 
one may use the new precision data to distinguish between the fundamental description in Herwiri1.031 and the
ad hocly hard intrinsic $p_T$ in Herwig6.5 by comparing the data to the predictions of the detailed line shape and of the more finely binned $p_T$ spectra --
see Figs.~3 and 4 in the last two papers in Refs.~\cite{herwiri1}\footnote{The discriminating power among the attendant theoretical predictions
of $p_T$ spectra in single $Z/\gamma^*$ production at the LHC is manifest in Refs.~\cite{pt-comp-LHC}-- the last paper in Refs.~\cite{herwiri1} provides more discussion on this point.}.
We await the availability of the new precision data accordingly.
\par
In closing, two of us (B.F.L.W., S.A.Y.)
thank Prof. Ignatios Antoniadis for the support and kind 
hospitality of the CERN TH Unit while part of this work was completed.
\par












\end{document}